\documentclass[sn-mathphys]{sn-jnl} 

\usepackage{lmodern}
\usepackage{enumitem}
\usepackage{amsmath,amssymb,bm}
\usepackage{xcolor}
\usepackage{hyperref} %
\usepackage{academicons} %
\usepackage{fontawesome5} %

\begin{document}

\author*[1,2]{\fnm{Maosheng} \sur{He}}\email{hmq512@gmail.com}
\author[1,2]{\fnm{Quanhan} \sur{Li}}
\author[3]{\fnm{Shun-Rong} \sur{Zhang}}
\author[4]{\fnm{Jeffrey M.} \sur{Forbes}}%
\author[5]{\fnm{Jiuhou} \sur{Lei}}
\author[6]{\fnm{Libo} \sur{Liu}}
\author[1]{\fnm{Jiankui} \sur{Shi}}
\author*[1]{\fnm{Chi} \sur{Wang}}\email{cw@spaceweather.ac.cn}

\affil*[1]{\orgdiv{Key Laboratory of Solar Activity and Space Weather}, \orgname{National Space Science Center, Chinese Academy of Sciences}, \orgaddress{\city{Beijing}, \country{China}}}
\affil[2]{\orgdiv{Hainan National Field Science Observation and Research Observatory for Space Weather}, \orgname{National Space Science Center, Chinese Academy of Sciences}, \orgaddress{\city{Beijing}, \country{China}}}
\affil[3]{\orgdiv{Haystack Observatory}, \orgname{Massachusetts Institute of Technology}, \orgaddress{\city{Westford}, \state{MA}, \country{USA}}}
\affil[4]{\orgdiv{Ann \& H. J. Smead Department of Aerospace Engineering Sciences}, \orgname{University of Colorado}, \orgaddress{\city{Boulder}, \country{USA}}}
\affil[5]{\orgdiv{Deep Space Exploration Laboratory/School of Earth and Space Sciences}, \orgname{University of Science and Technology of China}, \orgaddress{\city{Hefei}, \country{China}}}
\affil[6]{\orgdiv{Beijing National Observatory of Space Environment}, \orgname{Institute of Geology and Geophysics, Chinese Academy of Sciences}, \orgaddress{\city{Beijing}, \country{China}}}


\title{Reassessment of Ionospheric Responses to GRB~221009A: Disentangling Instrumental, Illumination and Geophysical Effects}

\abstract{
Gamma-ray bursts (GRBs) have long been proposed to perturb Earth's ionosphere, with occasional reports of disruptions in ultra- and extremely-low-frequency radio signals. The exceptionally bright GRB~221009A was recently claimed to induce multi-altitude ionospheric responses, including perturbations in satellite electric fields, regional total electron content (TEC), and the equatorial electrojet (EEJ). These claims have renewed interest in the potential near-Earth impacts of astrophysical transients. Here we perform an independent reassessment using expanded datasets spanning multiple altitudes. We find no coherent, burst-like TEC enhancement, show that the reported electric-field anomalies recur under specific illumination conditions each orbit, and demonstrate that the EEJ fluctuations preceded the burst and coincide with solar-wind variability. Together, these results indicate that the reported GRB-induced ionospheric responses are fully attributable to other natural geophysical processes and instrumental artefacts, thereby resolving a high-profile controversy and clarifying the true limits of GRBs’ ionospheric effects.
}


\maketitle

\section*{Introduction}

Gamma-ray bursts (GRBs) rank among the most energetic events in the universe, capable of releasing intense radiation across cosmological distances. Their potential to perturb Earth’s ionosphere has been intermittently suggested, primarily through observations of very- and extremely-low-frequency (VLF/ELF) radio variations~\cite{Fishman1988, Tanaka2011}. Yet, despite decades of GRB detections, ionospheric responses remain rare and controversial, with several studies reporting null results~\cite{Price2001}.

Proposed GRB-induced ionospheric effects typically involve the lower ionosphere, where ionization rates peak~\cite{Kasturirangan1976} and recombination processes dominate~\cite{SchunkNagy2009, He2009}. Consequently, transient perturbations are more readily detected at VLF/ELF frequencies, where reflections occur near these altitudes, rather than at higher-frequency.

Recently, Piersanti et~al.~\cite{Piersanti2023} reported potential upper ionospheric disturbances during the exceptionally luminous GRB~221009A, including electric-field perturbations at $\sim$500~km altitude, regional total electron content (TEC) enhancements, and equatorial electrojet (EEJ) fluctuations. If verified, these phenomena would represent a novel class of space-weather drivers beyond conventional solar or lower-atmospheric forcing, with implications for satellite navigation and communication systems.

Here we re-examine these claims using complementary datasets spanning multiple geophysical domains. Our analysis finds no measurable ionospheric response attributable to GRB~221009A, refuting the earlier interpretation.

\section*{Results}

We analyze global GNSS-derived TEC observations, a full year of electric-field data from the China Seismo-Electromagnetic Satellite (CSES), and EEJ proxies derived from ground magnetometers, together with concurrent solar-wind and geomagnetic parameters, to assess potential ionospheric responses to GRB~221009A.

\subsection*{Absence of Burst-Like TEC Enhancement}
\begin{figure}[htbp]
\centering
\includegraphics[width=0.99\textwidth]{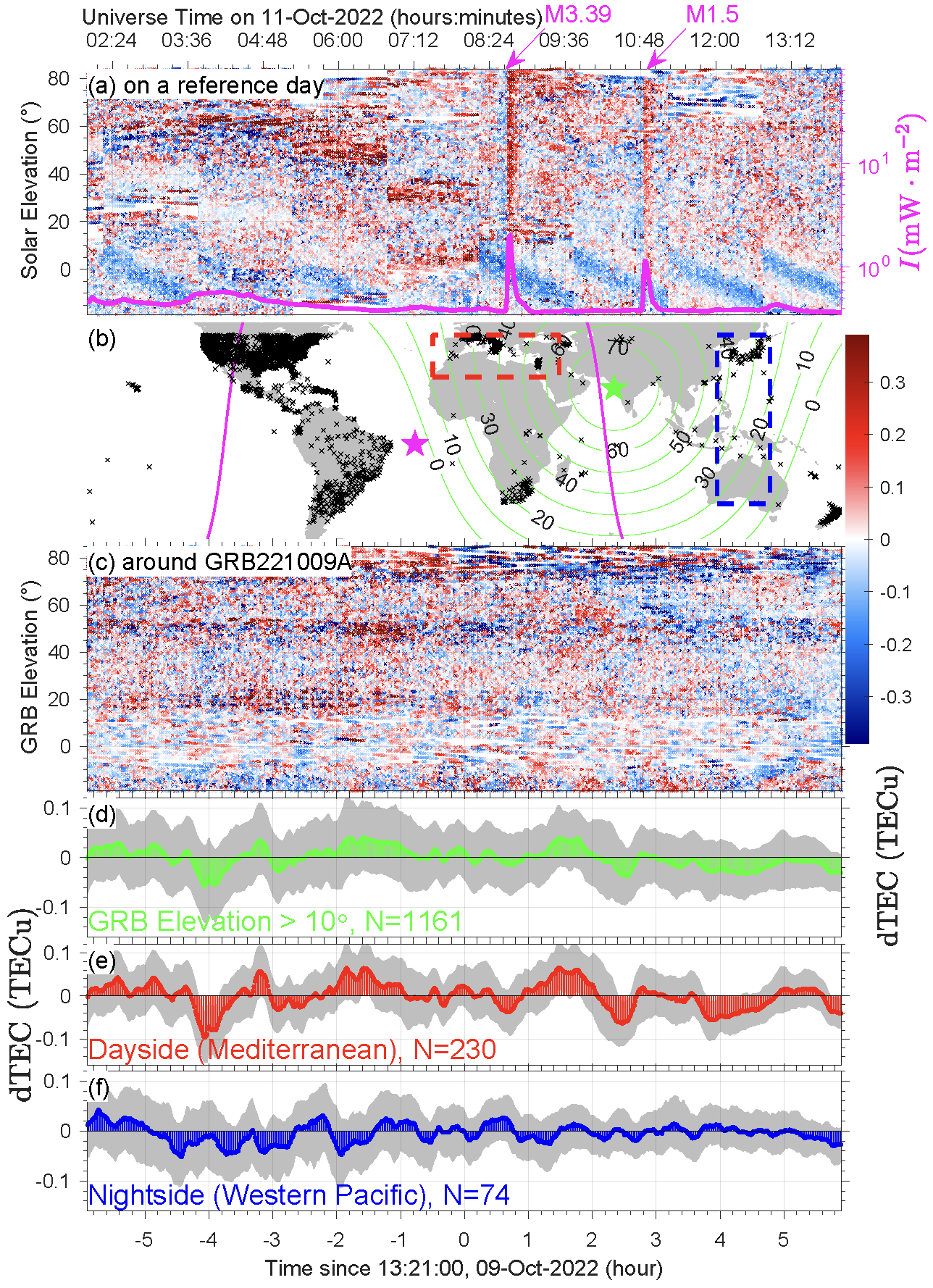}
\caption{High-pass–filtered TEC around two solar flares and GRB~221009A.
\textbf{a}, Filtered TEC on 11~October~2022 for two M-class flares (magenta arrows). Magenta line shows solar soft X-ray/EUV irradiance.
\textbf{b}, GNSS receiver distribution (black crosses). Green pentagon and lines indicate GRB zenith and elevation contour (zero marks illumination boundary); magenta symbols indicate solar zenith and terminator at GRB onset.
\textbf{c}, TEC versus GRB elevation within $\pm6$~h of the burst.
\textbf{d--f}, 2-min sliding medians (colored curves) and IQR (shading) for elevations $>10^\circ$ in the boxed regions in \textbf{b}.}
\label{fig:TEC_GRB}
\end{figure}

To evaluate whether GRB~221009A induced measurable ionospheric perturbations, we use solar flares as references, which typically produce short-lived, spatially coherent TEC enhancements. For example, two moderate M-class flares on 11~October~2022 generated increases below $0.35~\mathrm{TECu}$ over regions with solar elevations exceeding $10^\circ$ (Fig.~\ref{fig:TEC_GRB}a). TEC measurements (distribution shown in Fig.~\ref{fig:TEC_GRB}b) were high-pass filtered to remove variations on timescales longer than 30~min (Methods), allowing isolation of brief responses.

By contrast, TEC during GRB~221009A (Fig.~\ref{fig:TEC_GRB}c) exhibits only stochastic fluctuations, with no spatially coherent enhancement above $10^\circ$ GRB elevation. Analysis of 1,161 receivers within the illuminated hemisphere using a 2-min sliding median (Fig.~\ref{fig:TEC_GRB}d) reveals fluctuations below 0.03~$\mathrm{TECu}$ within $\pm1~h$ of the burst, far smaller than the typical interquartile range (IQR $>$0.12~$\mathrm{TECu}$). Subsets on both the dayside and nightside (Fig.~\ref{fig:TEC_GRB}e,f) show similarly negligible variability. These results indicate that GRB~221009A produced no detectable burst-like TEC response.

\subsection*{No Distinct Electric-Field Perturbations Attributable to GRB~221009A}

\begin{figure}[htbp]
\centering
\includegraphics[width=0.99\textwidth]{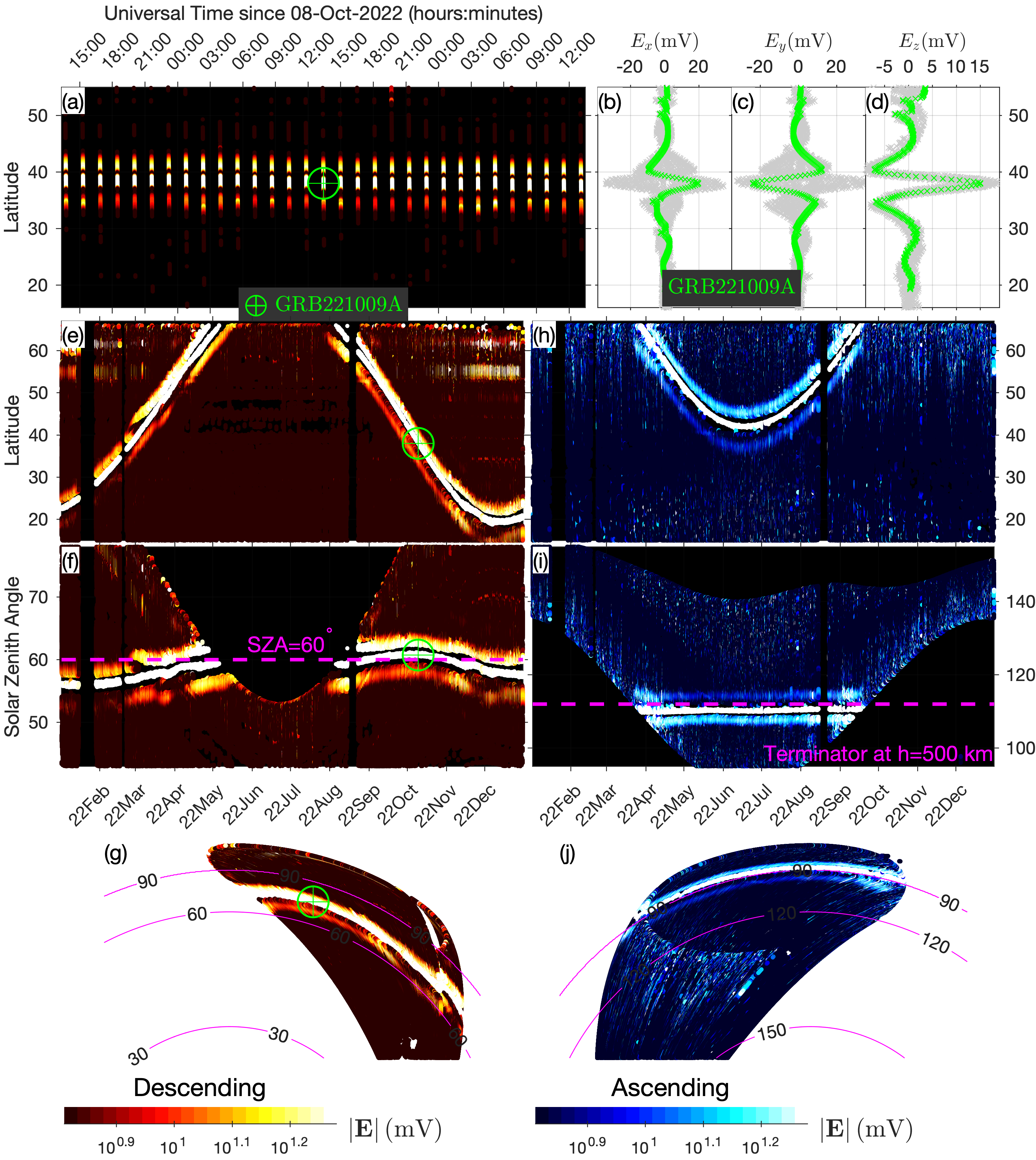}
\caption{CSES electric-field measurements within $\pm24$~h of GRB~221009A and throughout 2022.
\textbf{a}, Total electric-field amplitude for descending passes. Vertical stripes mark successive 95-min orbits; green circled plus ($\oplus$) indicates satellite position at burst onset.
\textbf{b--d}, Zonal ($E_x$), meridional ($E_y$), and vertical ($E_z$) components corresponding to \textbf{a}; green crosses indicate measurements within $\pm5$~min of onset.
\textbf{e--g}, Year-long total amplitude projected onto date-latitude, date-SZA, and GSE $y$-$z$ planes. Magenta lines show SZA isolines.
\textbf{h--j}, Same projections as textbf{e--g} but for ascending passes. Recurring perturbations correspond to illumination geometry rather than GRB effects.}
\label{fig:efield_combined}
\end{figure}

After removal of large-scale geomagnetic contributions (Methods), the CSES-measured electric field shows a perturbation in both its total amplitude and components around the time of GRB~221009A (Fig.~\ref{fig:efield_combined}a--d). However, this feature is not unique to the burst period—it recurs every orbit near 40$^\circ$N  and exhibits no anomaly compared with other orbits within the adjacent 48-hour window (Fig.~\ref{fig:efield_combined}a--d).  The recurrent perturbations  persist throughout 2022 (Fig.~\ref{fig:efield_combined}e--g) and migrate seasonally with solar zenith angle (SZA) boundaries $\mathrm{SZA}\approx60^\circ$, except for a gap in May--July attributable to orbital coverage limitations. CSES, in a sun-synchronous orbit at 507~km altitude with 97.4$^\circ$ inclination, completes 15--16 orbits per day, nominally sampling $\pm65^\circ$ latitude (extending to $\sim70^\circ$N in 2022)~\cite{Shen2018}. Gaps occur when the $\mathrm{SZA}\approx60^\circ$ isoline moves poleward beyond the sampled region (Fig.~\ref{fig:efield_combined}f).

The recurrent pattern, tied to illumination geometry rather than the GRB timing, indicates an instrumental or environmental origin.

\subsection*{EEJ Fluctuations Preceding GRB~221009A}
\begin{figure}[htbp]
\centering
\includegraphics[width=0.99\textwidth]{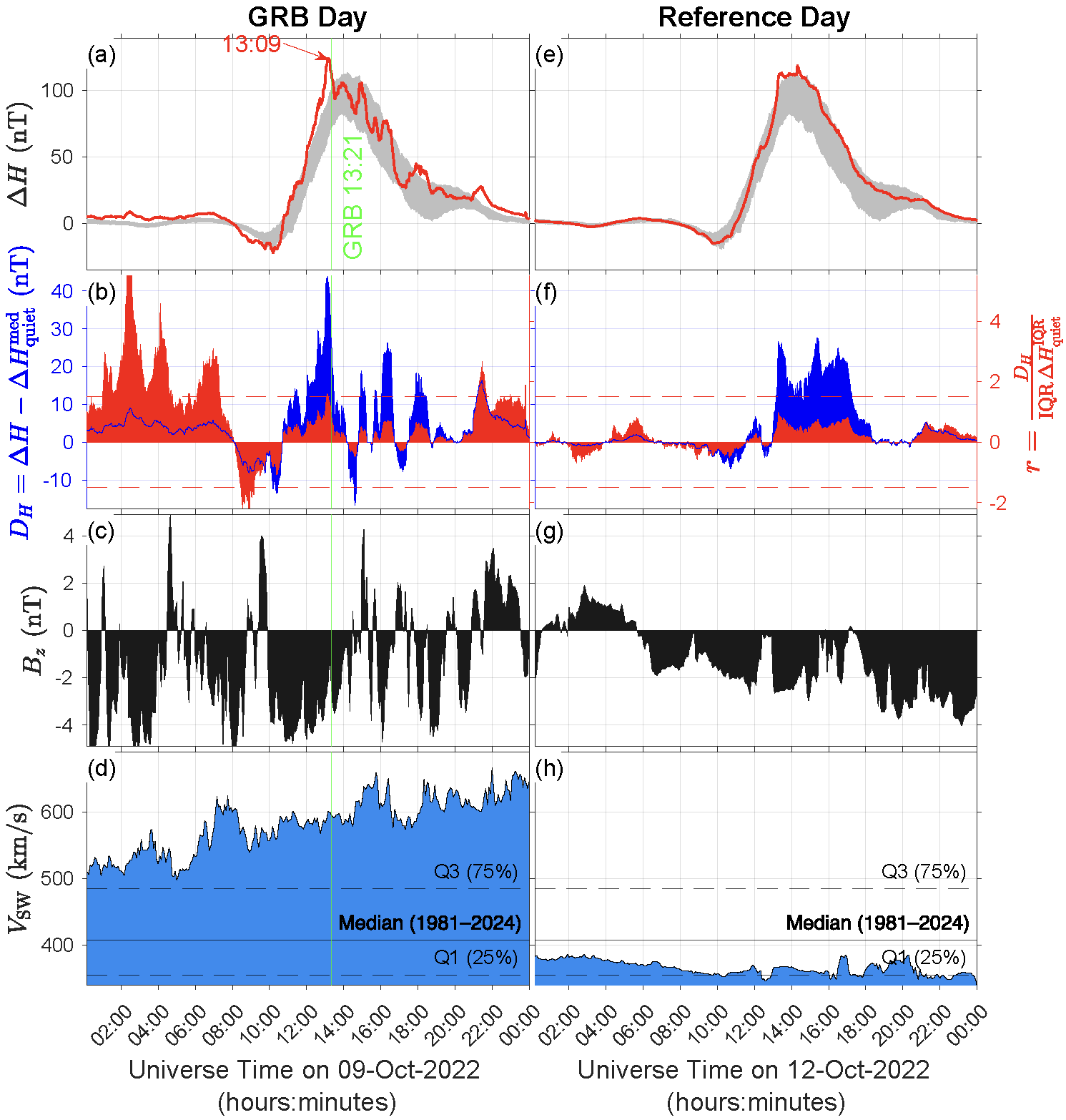}
\caption{EEJ proxies and solar-wind conditions on the GRB day versus a reference day.
\textbf{a}, EEJ proxy ($\Delta H$) on the GRB day. Shaded band indicates 30-day quiet-time IQR; red arrow marks peak $\sim$12~min before the burst.
\textbf{b}, Deviation from quiet-time median (blue) and its ratio to IQR (red), with $\pm150\%$ thresholds (dashed).
\textbf{c}, IMF $B_z$.
\textbf{d}, Solar-wind speed $V_\mathrm{SW}$ with climatological median and IQR (black).
\textbf{e--h}, Corresponding quantities on 12~October~2022. No GRB-related signatures are evident.}
\label{fig:EJJ_SW}
\end{figure}

During the GRB, the EEJ proxy ($\Delta H$) deviated from its quiet-time range (Fig.~\ref{fig:EJJ_SW}a). However, this deviation began hours before the burst and peaked about 12~min prior to the burst onset. The pre-burst peak becomes clearer after subtracting the quiet-time baseline ($D_H$ in Fig.~\ref{fig:EJJ_SW}b; see Methods). Normalized deviations ($r$ in Fig.~\ref{fig:EJJ_SW}b) remained below the 150\% IQR threshold afterward, indicating no significant post-burst anomaly. Therefore, the observed EEJ variability is unrelated to GRB~221009A.

\section*{Discussion}

All three diagnostic datasets—TEC, satellite electric-field, and EEJ—show no evidence of GRB-related effects, contradicting existing reports. In this section, we examine these discrepancies and show that the previously reported observations are fully attributable to known instrumental or geophysical processes rather than to GRB~221009A itself.

\subsection*{Discrepancy with Previously Reported TEC Enhancements}

Our TEC analysis reveals no enhancement attributable to GRB~221009A, directly challenging the previously reported $\sim$2~$\mathrm{TECu}$ increase~\cite{Piersanti2023}, which is comparable to the global effect of an X10-class solar flare~\cite{Zhang2019TID,Sarp2024}. This discrepancy is not due to spatial coverage—the Mediterranean sector analyzed in \cite{Piersanti2023} largely overlaps with our study region (red box in Fig.~\ref{fig:EJJ_SW}b, source of data in Fig.~\ref{fig:EJJ_SW}e)—but rather arises from how the “increase” was defined. In \cite{Piersanti2023}, the enhancement is defined as the mean TEC at 13:00–14:00 UT on 9 October relative to the same hour on the two surrounding days.

The day-to-day $\sim$2~TECu enhancement did not appear in our high-pass TEC evolution, suggesting that its characteristic timescale is longer than our 30-minute cut-off period. In particular, no abrupt increase occurred at the GRB onset, and thus the day-to-day enhancement cannot be attributed to the burst. Instead, it likely reflects background variability associated with elevated solar-wind speeds and IMF fluctuations preceding the event (Fig.~\ref{fig:EJJ_SW}c–d).

\subsection*{Recurrent Electric-Field Perturbations Linked to Illumination Geometry}
The recurrent CSES electric-field  perturbation coincides with transitions across illumination boundaries (SZA $\approx 60^\circ$), pointing to an instrumental origin. 

A plausible mechanism is the photoelectric effect on the electric-field detector (EFD) sensors: when conductive surfaces are illuminated, photoelectrons are emitted, and abrupt changes in illumination can drive transient currents. Differences in illumination timing or incidence angles between sensors—for example due to self-shadowing—can generate spurious electric-field pulses. Similar illumination-related artefacts have been reported for electric-field instruments on THEMIS (Time History of Events and Macroscale Interactions during Substorms), MMS (Magnetospheric Multiscale Mission), and Cluster~\cite{themis_efi_shadow,EDPGuide,eriksson2006}. 

These effects also explain another orbital periodicity: on ascending passes, perturbations align with $\mathrm{SZA} = 90^\circ$ boundaries (Fig.~\ref{fig:efield_combined}h--j), consistent with Earth's shadow crossings.

\subsection*{EEJ Variability Driven by Solar-Wind Forcing}
The EEJ was already deviating from quiet-time levels before GRB~221009A, ruling out any burst-related origin and contradicting earlier claims of GRB-induced low-frequency fluctuations.

Instead, these deviations are associated with elevated solar-wind speeds (up to $\sim$600~km~s$^{-1}$ with $\sim$50~km~s$^{-1}$ fluctuations) and rapid IMF $B_z$ reversals (Fig.~\ref{fig:EJJ_SW}c--d), likely injecting energy into the magnetosphere and producing the observed geomagnetic disturbances. By contrast, on the reference day (12 October; Fig.~\ref{fig:EJJ_SW}e--h), solar-wind conditions remained steady (350--400~km~s$^{-1}$, mostly southward $B_z$), in line with only minor EEJ deviations. The previously claimed EEJ response, inferred by comparison with the reference day, can thus be attributed to differing solar-wind conditions.
\section*{Conclusion}

Our multi-instrument reassessment finds no evidence that GRB~221009A produced any measurable ionospheric perturbation.The previously reported electric-field perturbations coincident with the burst recur every orbit and are attributable to self-shadowing instrumental artefactss, while the claimed TEC enhancement and EEJ fluctuation reflect day-to-day variations, with timings showing no association with the GRB. These   variations are consistent with solar-wind variability and associated geomagnetic disturbances. Together, our results highlight the importance of accounting for instrumental artefacts, illumination geometry, and space-weather conditions before attributing near-Earth perturbations to astrophysical transients, and offer broader methodological insight for future studies of astrophysical–geospace coupling.

\section*{Figures}\label{sec6}
\clearpage

\clearpage
\section*{Methods}

\subsection*{Global TEC Processing for Short-Term Variations}
Global Navigation Satellite System (GNSS) TEC data from 3,374 receivers between $45^\circ$N and $45^\circ$S were analyzed, excluding higher-latitude stations to minimize geomagnetic contamination. Observations with satellite elevation below $30^\circ$ were discarded. Time series were high-pass filtered to remove variations with periods $>30$~min, such as diurnal trends~\cite{Zhang2017Bow,Zhang2019TID}.

\subsection*{High-Pass Filtering of Electric-Field Data}
Raw EFD data were high-pass filtered to remove large-scale convective $\mathbf{v}\times\mathbf{B}$ fields. The cutoff period
\[
T_c = 3.7~\mathrm{min} = \frac{95~\mathrm{min}}{2 \times 13}
\]
corresponds to $\sim$1700~km along track, where 95~min/2 is half the orbital period and 13 is the relevant IGRF spherical-harmonic degree. A Butterworth filter preserves the passband smoothly. This approach avoids potential biases from geomagnetic models used in \cite{Piersanti2023}.

\subsection*{CSES E-Field Measurements and Illumination Effects}
The EFD employs a double-probe design, measuring potential differences between spherical sensors on a boom~\cite{Mozer2016}. Probe potentials depend on the balance between thermal-electron collection and photoelectron emission. Sunlight drives potentials positive via photoemission; in shadow, potentials drop. Transitions across illumination boundaries induce artificial potential differences, producing spurious electric-field signals~\cite{SarnoSmith2016}.

\subsection*{EEJ Proxy Derivation and Quiet-Time Normalization}
Magnetometer data from Tatuoca ($1.205^\circ$~S, $311.487^\circ$~E) near the magnetic equator and San Juan ($18.11^\circ$~N, $293.85^\circ$~E) outside the EEJ belt were used to estimate EEJ intensity~\cite{Yamazaki2017}. One-second data were resampled to 1~min where necessary. Long-term and non-cyclic variations were removed by interpolating baselines between quiet nighttime intervals, defined as the four-hour window centered on local midnight with mean the symmetric disturbance index in H-component (SYM-H) $>-20$~nT~\cite{Siddiqui2015}. 

The EEJ proxy $\Delta H$ was computed as the baseline-corrected north-component difference between stations. From the 30-day quiet interval surrounding the GRB, the median
\[
\Delta H_{\mathrm{quiet}}^{\mathrm{med}}
\]
and IQR
\[
\Delta H_{\mathrm{quiet}}^{\mathrm{IQR}} = \Delta H_{\mathrm{quiet}}^{\mathrm{Q3}} - \Delta H_{\mathrm{quiet}}^{\mathrm{Q1}}
\]
were obtained, where Q1 and Q3 denote the first and third quartiles. Deviations from quiet conditions were defined as
\[
D_H = \Delta H - \Delta H_{\mathrm{quiet}}^{\mathrm{med}}, 
\quad
r = \frac{D_H}{\Delta H_{\mathrm{quiet}}^{\mathrm{IQR}}},
\]
quantifying perturbation magnitudes relative to typical quiet-time variability.

\backmatter
\section*{Acknowledgments}
This work is supported by the Strategic Priority Research Program of the Chinese Academy of Sciences (XDA0470301), the National Natural Science Foundation of China (42188101) and the Chinese Meridian Project.

\section*{Data Availability}

The datasets used in this study are publicly available from the following sources:

\begin{itemize}
\item Level-2 2022 ULF electric field waveforms and power spectra from the EFD instrument onboard CSES-01: \url{http://www.leos.ac.cn/#/dataService/dataBrowsingList} (accessed 1 June 2025).
\item GNSS TEC data from the Massachusetts Institute of Technology: \url{http://www.openmadrigal.org} (accessed 8 August 2025).
\item Ground magnetometer data from INTERMAGNET: \url{https://www.intermagnet.org} (accessed 20 September 2025).
\item Solar wind parameters and the SYM-H index from NASA OMNIWeb: \url{https://omniweb.gsfc.nasa.gov/} (accessed 20 July 2025).
\item Full-disk soft X-ray and EUV solar irradiance from the Extreme Ultraviolet Spectrophotometer (ESP) onboard EVE/SDO: \href{https://lasp.colorado.edu/eve/data_access/eve-one-minute-averages/index.html}{LASP EVE data portal} (accessed 1 October 2022).
\end{itemize}

\section*{Code Availability}
MATLAB codes used to generate results and figures are available from the corresponding author upon reasonable request.

\end{document}